\documentclass[twocolumn,aps,prl]{revtex4}

\usepackage{graphicx}
\usepackage{amsmath,amssymb}
\usepackage{textcomp}
\usepackage{bibunits}

\newcommand{\vc}[1]{\ensuremath{\mathbf{#1}}}

\newcommand{\ket}[1]{\ensuremath{\left|  #1 \right\rangle}}

\newcommand{\aver}[1]{\ensuremath{\langle {#1} \rangle}}
\newcommand{\eaver}[1]{\ensuremath{\langle {#1} \rangle_e}}
\newcommand{\abs}[1]{\ensuremath{\left| {#1} \right|}}
\newcommand{\sdev}[1]{\ensuremath{\Delta #1}}

\newcommand{\var}[1]{\ensuremath{\Delta{#1^2}}}

\newcommand{\meanS}{\ensuremath{|\langle\vc{S}\rangle|}}
\newcommand{\Rb}{\ensuremath{{}^{87}\mathrm{Rb}}}

\newcommand{\zetaW}{\ensuremath{\zeta_\mathrm{m}}}
\newcommand{\zetaKU}{\ensuremath{\zeta_\mathrm{e}}}

\newcommand{\Sin}{\ensuremath{S_{\mathrm{in}}}}
\newcommand{\CSSvar}{\ensuremath{\var{S_z}_{\mathrm{CSS}}}}
\newcommand{\Techvar}{\ensuremath{\var{S_z}_\mathrm{tech}}}
\newcommand{\Measvar}{\ensuremath{\var{S_z}_\mathrm{meas}}}
\newcommand{\Prepvar}{\ensuremath{\var{S_z}_\mathrm{prep}}}
\newcommand{\F}{\ensuremath{\mathcal{F}}}
\newcommand{\Var}[1]{\ensuremath{\mathrm{Var}(#1)}}
\newcommand{\Mt}{\ensuremath{\tilde{M}}}
\newcommand{\Condvar}{\ensuremath{[\var{S_z}]_{M_1}}}
\newcommand{\Coop}{\ensuremath{N_0\eta_\mathrm{eff}}}
\newcommand{\Psc}{\ensuremath{P_\mathrm{sc}}}
\newcommand{\flipin}{\ensuremath{P_{\Delta F}}}
\newcommand{\flipout}{\ensuremath{P_{\Delta F \Delta m_F}}}
\newcommand{\shift}{\ensuremath{P_{\Delta m_F}}}
\newcommand{\PRam}{\ensuremath{P_\mathrm{Ram}}}
\newcommand{\phieff}{\ensuremath{\phi_\mathrm{eff}}}
\newcommand{\Cov}[2]{\ensuremath{\mathrm{Cov}(#1,#2)}}

\newcommand{\Szf}{\ensuremath{{S_{zf}}}}

\newcommand{\tM}{\ensuremath{\tilde{M}}}

\begin{document}
\begin{bibunit}

\title{States of an Ensemble of Two-Level Atoms with Reduced Quantum Uncertainty}

\author{Monika H. Schleier-Smith}
\affiliation{
Department of Physics, MIT-Harvard Center for Ultracold Atoms,
and Research Laboratory of Electronics, Massachusetts Institute of Technology,
Cambridge, Massachusetts 02139, USA}

\author{Ian D. Leroux}
\affiliation{
Department of Physics, MIT-Harvard Center for Ultracold Atoms,
and Research Laboratory of Electronics, Massachusetts Institute of Technology,
Cambridge, Massachusetts 02139, USA}

\author{Vladan Vuleti\'{c}}
\affiliation{
Department of Physics, MIT-Harvard Center for Ultracold Atoms,
and Research Laboratory of Electronics, Massachusetts Institute of Technology,
Cambridge, Massachusetts 02139, USA}

\date{\today}
\begin{abstract}
We generate entangled states of an ensemble of $5 \times 10^4$ ${}^{87}$Rb atoms by optical quantum nondemolition measurement.  The resonator-enhanced measurement leaves the atomic ensemble, prepared in a superposition of hyperfine clock levels, in a squeezed spin state.  By comparing the resulting reduction of quantum projection noise (up to 8.8(8)~dB) with the concomitant reduction of coherence, we demonstrate a clock input state with spectroscopic sensitivity 3.0(8)~dB beyond the standard quantum limit.

\end{abstract}

\maketitle

Atomic clocks \cite{Santarelli99,Ludlow08,Takamoto05} and atom interferometers \cite{Durfee06} are reaching the standard quantum limit (SQL) of precision \cite{Wineland92,Wineland94,Santarelli99}, set by the quantum projection noise inherent in measurements on a collection of uncorrelated particles.  In the canonical Ramsey interferometer with $N_0$ particles, a quantum mechanical phase is converted into occupation probabilities for two states and read out as a population difference $N$ between them. Entanglement can reduce the projection noise $\Delta N$ by redistributing it to another variable that does not directly affect the experiment precision. The resulting ``squeezed spin state'' \cite{Kitagawa93,Meyer01,Leibfried04,Kuzmich98,Sorensen01,Bouchoule02,Madsen04,Hammerer04,Hald99,Andre04} can be used as an input state to an interferometer to overcome the SQL \cite{Wineland92,Wineland94,Meyer01,Leibfried04}.

Formally, the system can be described by an ensemble spin vector $\vc{S}=\sum \vc{s}_i$ that is the sum over the (pseudo-\nolinebreak) spins $\vc{s}_i$ of the individual (spin-1/2) particles \cite{Wineland92,Wineland94,Kitagawa93}. The ensemble spin $S$ with $\aver{\vc{S}^2}=S(S+1)$ can take on values in the range $0 \leq S \leq S_0$, where $S_0=N_0/2$. For a given $S$, the minimum variance $\var{S_z}$ of $S_z=N/2$ for an unentangled state is realized by the coherent spin state (CSS), and is given by $\CSSvar=S/2=\meanS/2$, where it is assumed that the mean ensemble spin vector $\aver{\vc{S}}$ lies in the $xy$-plane. A spin state can be defined as squeezed if it satisfies $\zetaKU \equiv 2 \var{S_z}/\meanS<1$ (entanglement criterion \cite{Kitagawa93,Sorensen01}), or $\zetaW \equiv2 \var{S_z} \Sin/\meanS^2 < 1$ (criterion for metrological gain \cite{Wineland92,Wineland94}, where $\Sin$ is the initial spin of the uncorrelated ensemble before the squeezing). $\zetaW^{-1}$ represents the increase in the squared signal-to-noise ratio $\meanS^2/\var{S_z}$ over the value $2 \Sin$ for the initial uncorrelated state. Since $\meanS \leq \Sin$, we have $\zetaKU \leq \zetaW$, i.e. metrological gain guarantees entanglement.

The process utilized for spin squeezing can reduce $\meanS$ below the initial spin $\Sin$ before the squeezing, thereby reducing the minimum variance $\var{S_z}$ that is consistent with an unentangled state \cite{Sorensen01}. Therefore, measurements of both spin noise $\sdev{S_z}$ and average spin length after squeezing $|\aver{\vc{S}}|$ are necessary to verify spin squeezing or quantify metrological gain. While reduction of spin noise alone has sometimes been referred to as ``spin squeezing'' \cite{Kuzmich00,Takano09} or ``number squeezing" \cite{Gerbier06,Jo07}, we take spin squeezing to require at least demonstrated entanglement, $\zetaKU<1$, although we are primarily interested in metrological gain, $\zetaW<1$.


\begin{figure}
\includegraphics[width = 3.4 in]{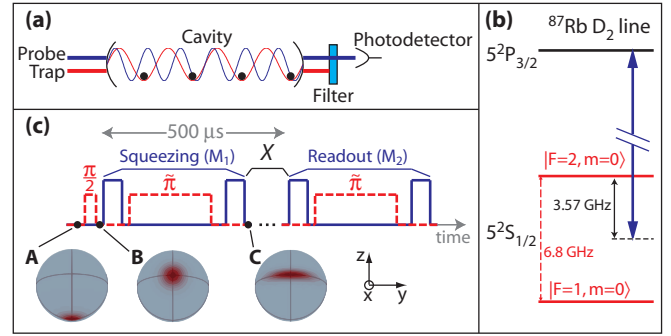}
\caption{\label{fig:1setup} (a) Experimental setup. (b) Atomic level structure. (c) Experimental sequence.  Timing of probe pulses (solid line) and microwave pulses (dashed line) in preparation and readout of a squeezed state.  $\tilde{\pi}$ designates a composite $\pi$ pulse \cite{EPAPS}.  Various procedures are inserted at $X$, as described in the text, to measure the CSS variance, measure the noise of a spin component other than $S_z$, or operate a clock. \textbf{A}-\textbf{C} illustrate semiclassical probability distribution functions for the Gaussian states discussed in the text.}
\end{figure}

Spin noise has been modified by atomic collisions \cite{Orzel01,Gerbier06,Jo07} and by absorption of squeezed light \cite{Hald99}.  In dilute atomic systems, quantum nondemolition (QND) measurements with light \cite{Kuzmich98,Sorensen01,Bouchoule02,Madsen04,Kuzmich00,Teper08,Takano09} have reduced the projection noise of rotating \cite{Kuzmich00} and stationary \cite{Takano09} spins.  Spin squeezing has been achieved with two ions \cite{Meyer01}, and spectroscopic sensitivity further improved with a maximally entangled state of three ions \cite{Leibfried04}. Recently, spin squeezing with a Bose-Einstein condensate (BEC) in a multiple-well potential has been reported \cite{Esteve08}. Demonstrated metrological gains over the SQL include $\zetaW^{-1}=3.2(1)$~dB in the three-ion system \cite{Leibfried04}; $\zetaW^{-1}\sim 4$~dB by light-induced squeezing within individual atoms of large spin $s=3$ \cite{Chaudhury07}; and $\zetaW^{-1}=3.8(4)~$dB for the BEC \cite{Esteve08}.

In this Letter, we demonstrate the generation of squeezed spin states of $5 \times 10^4$ trapped \Rb\ atoms on an atomic-clock transition by resonator-aided QND measurement with a far-detuned light field, as proposed by Kuzmich, Bigelow, and Mandel \cite{Kuzmich98}.  We verify the entanglement by comparing the observed reduction in projection noise below that of a coherent spin state (up to 8.8(8)~dB) with the accompanying reduction in clock signal, and achieve a $3.0(8)$~dB improvement in precision over the SQL.

The light-induced spin squeezing presented here requires strong ensemble-light coupling \cite{Kuzmich98,Bouchoule02,Madsen04,Hammerer04} (large collective cooperativity \cite{EPAPS}). This is achieved by means of a near-confocal optical resonator with, at the $2\pi/k=780~\mathrm{nm}$ wavelength of the probe light, a finesse $\F=5.6(2) \times 10^3$, a linewidth $\kappa = 2\pi\times 1.01(3)\mathrm{MHz}$, and a mode waist $w = 56.9(4)~\hbox{\textmu}\mathrm{m}$ at the atoms' position, corresponding to a maximal single-atom cooperativity $\eta_0=24\F/(\pi k^2w^2)=0.203(7)$ \cite{EPAPS}.  Our experiments are performed on an ensemble containing up to $N_\mathrm{a}=5\times 10^4$ laser-cooled \Rb\ atoms optically trapped inside the resonator in a standing wave of 851-nm light (Fig. \ref{fig:1setup}).

One resonator mode is tuned $3.57(1)~\mathrm{GHz}$ to the blue of the $\ket{5^2S_{1/2},F=2}\rightarrow\ket{5^2P_{3/2},F'=3}$ transition in \Rb, such that the atomic index of refraction results in a mode frequency shift $\omega$ that is proportional to the population difference $N=N_2-N_1$ between the hyperfine clock states $\ket{1}=\ket{5^2S_{1/2},F=1,m_F=0}$ and $\ket{2}=\ket{5^2S_{1/2},F=2,m_F=0}$. The transmission of a probe laser tuned to the slope of this mode thus directly measures $S_z=N/2$, and is insensitive to total atom number (Fig. \ref{fig:1setup}). The atom-resonator coupling also gives rise to a differential light shift between the clock states, which we use to verify experimentally the coupling strength calculated from first principles from spectroscopically determined resonator parameters.  We measure a phase shift of $250(20)~\hbox{\textmu}\mathrm{rad}$ per  transmitted photon for a maximally coupled atom (on the resonator axis at an antinode of the probe standing wave), in excellent agreement with the calculated value $253(8)~\hbox{\textmu}\mathrm{rad}$ \cite{EPAPS}.

To account for the spatial variation in coupling between standing-wave probe light and atoms, we define an effective atom number $N_0=(\eaver{\eta}^2/\eaver{\eta^2})N_\mathrm{a} \approx 0.66 N_\mathrm{a}$, where the single-atom cooperativity $\eta$, proportional to the local intensity of probe light, is averaged over the ensemble containing $N_\mathrm{a}$ atoms \cite{EPAPS}.  The definition is chosen so that the projection noise variance of the effective atom number measured via the mode shift $\omega\propto N_a\eaver{\eta}$ satisfies the usual relation $\var{N_0}=N_0$.  This avoids carrying near-unity factors through the equations and allows direct comparison to a spatially uniform system of collective cooperativity $N_0\eta_\mathrm{eff}$, where $\eta_\mathrm{eff} = (2/3)\eaver{\eta^2}/\eaver{\eta}=0.47(1)\eta_0$, taking into account the oscillator strength 2/3 of the $D_2$ line and the measured rms transverse cloud radius of $8.1(8)~\hbox{\textmu}\mathrm{m} \ll w$.  The mode frequency shift per effective atom of population difference $N$ between the clock states is $d\omega/dN=4.5(2)\times 10^{-5}~\kappa$ \cite{EPAPS}.

To quantify spin squeezing we need to measure $\var{S_z}$ and $\meanS$. The latter can be obtained from the observed contrast $\mathcal{C}$ of Rabi oscillations  as $\meanS=\mathcal{C} S_0$, where the maximum spin $S_0=N_0/2$ is measured by optically pumping the atoms between the two hyperfine states $F=1,2$. For large $S_0$ the cavity shift $\omega$ exceeds $\kappa$ ($\omega \le 1.8 \kappa$), which we take into account by correcting for the (accurately measured) Lorentzian lineshape of the resonator.  To verify the atom numbers $2S_0$ thus obtained, we have also directly measured the cavity mode frequency shift $\omega \propto S_z$, finding agreement to within 2(4)\% \cite{EPAPS}.  $\var{S_z}$ is obtained from transmission measurements that always remain in the linear regime, with $2 |\Delta S_z| d\omega/dN\le 0.01\kappa$.

The probe laser is frequency-stabilized to a far detuned, negligibly shifted mode \cite{EPAPS}. Each measurement of $S_z$ employs two probe light pulses of duration $T = 50~\hbox{\textmu}\mathrm{s} \gg \kappa^{-1}= 158~\mathrm{ns}$ separated by a $280~\hbox{\textmu}\mathrm{s}$ delay, during which we apply a microwave $\pi$ pulse sequence \cite{EPAPS} to suppress inhomogeneous light shifts (spin echo sequence).  Each probe light pulse contains $10^5$ to $10^6$ photons which, after traversing the resonator, are detected with an overall quantum efficiency $Q_e=0.43(4)$.  From the detected photon numbers in the two pulses, we deduce two cavity shifts $\omega_\pm$ whose difference constitutes a single measurement $M$ of $S_z = (\omega_+-\omega_-)/(4d\omega/dN)$.  In a typical experiment (Fig. 1(c)), after initializing the ensemble spin state by optical pumping into $\ket{1}$ (\textbf{A}) and applying a $\pi/2$ microwave pulse to rotate the CSS into an equal superposition of $\ket{1}$ and $\ket{2}$ (\textbf{B}), we perform two measurements $M_1$ and $M_2$ to induce and verify conditional spin squeezing.  We quantify spin noise $\sdev{S_z}$ by extracting variances from 100 repetitions of such a sequence.

We determine the CSS projection noise level $\CSSvar=N_0/4$ from the measured atom number $N_0$ and verify it \cite{Wineland92,Wineland94,Hald99,Kuzmich00} either by evaluating the variance $\Var{M_1}$ of the set of single measurements $M_1$; or by inserting between two measurements $\Mt_1$ and $\Mt_2$ a second CSS preparation, consisting of optical pumping into state $\ket{1}$ and a $\pi/2$ pulse, and evaluating $\Var{\Mt_1-\Mt_2}/2$. Fig.~\ref{fig:2SpinMeasurementVariance} shows the dependence of the corresponding quantities in atom number units, $y_1=4\Var{M_1}$ (open triangles) or $y_2=2\Var{\Mt_1-\Mt_2}$ (open circles), on $N_0$. The contribution of CSS projection noise scales as $\CSSvar \propto N_0$, while atom-number-dependent technical noise, e.g. due to microwave power fluctuations or any sensitivity to atom number fluctuations, generically scales as $\Techvar \propto N_0^2$. A quadratic fit  $y_{1,2}=a_0+a_1 N_0+a_2 N_0^2$ yields $a_1=1.3(1)$ and $a_2=1(2)\times 10^{-6}$ (not shown in Fig. \ref{fig:2SpinMeasurementVariance}), but the data are also well fit by setting $a_1=1$, as required by independently measured cavity and atomic properties with no free parameters \cite{EPAPS}, and allowing a small technical noise contribution $a_2 N_0^2 < N_0$ with $a_2=9(3)\times10^{-6}$ (solid curve). Slow drifts in microwave power of $0.4\%$ over the set of measurements could account for the technical noise of $y_1$, which vanishes if the data are analyzed by comparing only adjacent cycles of the experiment \cite{EPAPS}.  Our ability to prepare an unentangled state close to a CSS---with $S_z$ variance $\Prepvar\sim 1.3S_0/2$ for our largest atom number---is not a prerequisite for spin squeezing but does provide independent confirmation of the CSS reference level for spin noise measurements. We emphasize that, in quantifying spin squeezing below, we conservatively normalize to the CSS noise $4\CSSvar=N_0$ as obtained from our cavity parameters (dashed line), not to the 30\% larger slope of the unconstrained quadratic fit to $y_{1,2}$.

\begin{figure}
\includegraphics[width = 3.4 in]{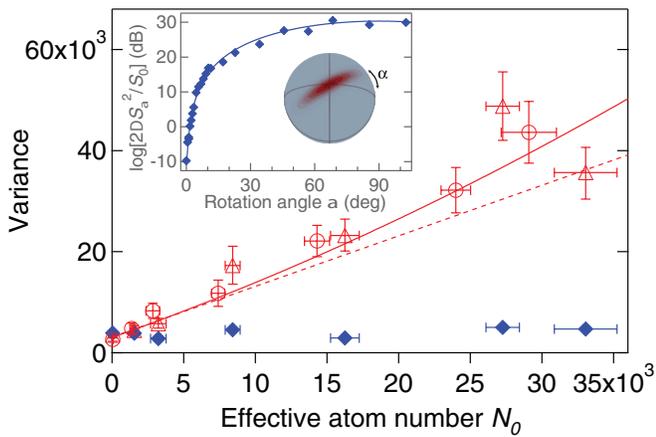}
\caption{\label{fig:2SpinMeasurementVariance} Spin noise measurements (see text): $y_1=4\Var{M_1}$ for a single CSS preparation (open triangles), $y_2=2\Var{\Mt_1-\Mt_2}$ for two independently prepared CSSs (open circles), and $4\Measvar=2\Var{M_1-M_2}$ for two measurements after a single CSS preparation (solid diamonds), all in units of atom number.  Vertical error bars are statistical; horizontal error bars indicate standard deviation of measured atom numbers. The solid (dashed) line corresponds to $a_1=1,a_2=9 \times 10^{-6}$ ($a_2=0$). Inset: Variance $\Delta S_\alpha^2$ of $S_z$ after rotating a squeezed state by an angle $\alpha$ about $\aver{\vc{S}}$, with parameter-free theory curve.}
\end{figure}

To prepare a state with (conditionally) reduced $\var{S_z}$ (Fig. 1(c)\textbf{C}), we simply measure $S_z$ for a CSS on the $x$-axis with a photon number $p\approx 5\times 10^5$ sufficiently large to resolve $S_z$ beyond the CSS variance.  Each such measurement $M_1$ yields a value of $S_z$ that is random but known, as verified by a readout measurement $M_2$.  We plot $2\Var{M_1-M_2}$ vs. atom number $N_0$ in Fig.~\ref{fig:2SpinMeasurementVariance} (solid diamonds), finding it a factor of 2 above the photocurrent noise level, with very weak dependence on atom number, and well below the CSS level.

In principle it is possible for the value of $S_z$ at the end of the measurement to differ from the average value of $S_z$ during the measurement. Besides the far-detuned locking light whose effect on $S_z$ is negligible, only spin-echo microwave composite $\pi$ pulses, whose fidelity was separately measured to be 98(1)\%, and probe light are applied during $M_1$. The probe light can only change $S_z$ through free-space scattering, which adds at most 3.1(3)\% of CSS projection noise at $p=5\times 10^5$ \cite{Madsen04,EPAPS}. Thus, while the added noise is negligible compared to the CSS level, it can explain part of the small remaining variance of $M_1-M_2$.

Provided $M_1$ does not alter the state appreciably, and the measurements $M_1,M_2$ are identical and uncorrelated \cite{EPAPS}, $\Measvar \equiv \Var{M_1-M_2}$/2 represents the uncertainty of any single such measurement.  The conditional variance of the state after measurement $M_1$ can then be shown to be $\Condvar=\Prepvar\Measvar/(\Prepvar+\Measvar)$ \cite{EPAPS}. When no new information is gained in measurement $M_1$ ($\Measvar \gg \Prepvar$), the variance is that of the state preparation process, $\Prepvar\equiv \Var{M_1}-\Measvar$ (close to, but above, the CSS value), while information gained reduces the variance, ultimately to the measurement variance $\Measvar$ of $M_1$.  At $N_0 = 3.3(2)\times 10^4$ and $p=6 \times 10^5$, we observe a normalized spin noise $\sigma^2 \equiv\Condvar/\CSSvar=-9.1(8)$~dB (see Fig. \ref{fig:3SqueezingContrast}); a slight correction for the effect of photon scattering \cite{EPAPS} yields $\sigma^2 =-8.8(8)$~dB.

The reduction of $\Condvar$ below the CSS value $\CSSvar$ is accompanied by a substantial increase in $\var{S_y}$ because the differential light shift of the atomic levels, corresponding to a rotation of the Bloch vector about the $z$ axis, depends on the intracavity intensity, which in turn depends on $S_z$. To observe the antisqueezing, we apply a microwave pulse after the squeezing measurement (at $X$ in Fig. 1(c)) to rotate the spin state by a variable angle $\alpha$ about $\aver{\vc{S}}$ before reading out $S_z$.  The variance $\var{S_\alpha}$ of $S_z$ in the rotated state, displayed in the inset to Fig.~\ref{fig:2SpinMeasurementVariance}, is a sinusoid that is well described with no free parameters by our model of the ensemble-cavity interaction \cite{EPAPS}.

To verify spin squeezing, we also need to measure $\meanS$, observable as the interference contrast $\mathcal{C} = \meanS/S_0$ of Rabi oscillations induced between measurements $M_1$ and $M_2$. Fig. \ref{fig:3SqueezingContrast} shows $\mathcal{C}$ as a function of photon number $p$ used in the state-preparation measurement at $N_0 = 4.0(1)\times 10^3$, and we have verified that the contrast $\mathcal{C}$ is independent of atom number \cite{EPAPS}.  Both normalized spin noise $\sigma^2$ and $\mathcal{C}$ can be fit by simple models (dashed and dotted curves) \cite{EPAPS}.  From these two measurements, we deduce the metrological squeezing parameter $\zetaW$ (solid triangles and solid curve).  For $p=3 \times 10^5$, we achieve $\zetaW^{-1}=\mathcal{C}^2/(\sigma^2 \mathcal{C}_\mathrm{in})= 3.0(8)$~dB of metrological gain (and an inverse entanglement parameter $\zetaKU^{-1}=\mathcal{C}/\sigma^2= 4.2(8)$~dB, not shown).  The finite initial contrast $\mathcal{C}_{\mathrm{in}}=S_\mathrm{in}/S_0=0.7$ in the ensemble without squeezing is due to the resonator locking light, and can be improved by detuning this light further from atomic resonance.  The probe-induced contrast reduction probably arises from differential light shifts between the clock states that are imperfectly canceled by the spin echo technique because of atomic motion. In the absence of any technical noise, a fundamental limit to the spin squeezing, associated with photon scattering into free space, would be $\zetaW^{-1}\le\sqrt{(3/2)N_0\eta_\mathrm{eff}} \sim 18~\mathrm{dB}$ in our system with cooperativity $N_0\eta_\mathrm{eff}\sim 3100$ \cite{Madsen04,Hammerer04,EPAPS}.

\begin{figure}
\includegraphics[width = 3.4 in]{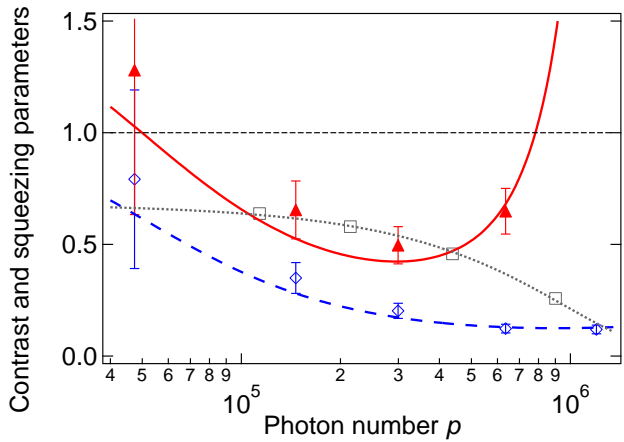}
\caption{\label{fig:3SqueezingContrast} Measured data with fits of simple models \cite{EPAPS} for normalized spin noise $\sigma^2=\Condvar/\var{S_z}_{\mathrm{CSS}}$ (open diamonds), contrast $\mathcal{C}$ (open squares), and metrological squeezing parameter $\zetaW$ (solid triangles).}
\end{figure}

For the data presented above, the readout quantifying the entanglement was completed $500~\hbox{\textmu}\mathrm{s}$ after preparation of the squeezed state.  We have further verified that the squeezing remains after a Ramsey clock sequence, in which two $\pi/2$ pulses about the $x$-axis, separated by a short ($70~\hbox{\textmu}\mathrm{s}$) precession time, are inserted at $X$ in Fig.~\ref{fig:1setup}c.  Such a clock can achieve precision below the SQL because the first of these $\pi/2$ rotations initiates it with a phase that is known, from the squeezing measurement, to better than the CSS uncertainty.

The phase coherence time of the unsqueezed CSS in our current trap is $10(2)~\mathrm{ms}$.  Both microwave and optical clocks with $\sim 1$~s coherence times have already been demonstrated with trapped atoms \cite{Harber02,Treutlein04,Takamoto05,Ludlow08,Ye08}.  Whether and to what degree the squeezing technique demonstrated here could benefit such clocks and other precision experiments \cite{Durfee06} will depend on the clock characteristics, noise sources \cite{Andre04}, and lifetime of the squeezed state. These questions, as well as possible systematic effects, need to be investigated in the future.

The group of E.~Polzik independently and simultaneously achieved results similar to ours in a Mach-Zehnder interferometer \cite{Appel09}.  We have recently demonstrated a new squeezing method using cavity feedback \cite{Leroux10}.

We thank J.K.~Thompson, M.D.~Lukin, D.~Stamper-Kurn, and E.~Polzik for
interesting discussions. This work was supported in part by the NSF, DARPA, and
the NSF Center for Ultracold Atoms.  M.~S. acknowledges support from the Hertz
Foundation and NSF.  I.~D.~L.  acknowledges support from NSERC.

\end{bibunit}

\begin{bibunit}
\newcounter{sec}
\newcounter{subsec}[sec]
\renewcommand{\thesec}{\roman{sec}}
\renewcommand{\thesubsec}{\Alph{subsec}}
\let\stdsection\section
\let\stdsubsection\subsection
\renewcommand{\section}[1]{\refstepcounter{sec}\stdsection{\thesec.\quad#1}}
\renewcommand{\subsection}[1]{\refstepcounter{subsec}\stdsubsection{\thesubsec.\quad#1}}

\setcounter{figure}{0}
\setcounter{page}{1}

\renewcommand{\thefigure}{A\arabic{figure}}
\renewcommand{\thetable}{A\arabic{table}}

\title{States of an Ensemble of Two-Level Atoms with Reduced Quantum Uncertainty:\\ Auxiliary Material}

\author{Monika H. Schleier-Smith}
\affiliation{Department of Physics, MIT-Harvard Center for Ultracold Atoms, and Research Laboratory of Electronics, Massachusetts Institute of Technology, Cambridge, Massachusetts 02139, USA}

\author{Ian D. Leroux} \affiliation{Department of Physics, MIT-Harvard Center for Ultracold Atoms, and Research Laboratory of Electronics, Massachusetts Institute of Technology, Cambridge, Massachusetts 02139, USA}

\author{Vladan Vuleti\'{c}}
\affiliation{Department of Physics, MIT-Harvard Center for Ultracold Atoms, and Research Laboratory of Electronics, Massachusetts Institute of Technology, Cambridge, Massachusetts 02139, USA}

\date{\today}

\maketitle

\section{Optical Resonator and Dipole Trap}\label{sec:loading}

The parameters of the near-confocal Fabry-P\'{e}rot resonator at the wavelengths of trap light (851 nm) and probe light (780 nm) are summarized in table \ref{tab:resonator}.  Both trap laser and probe laser are locked to the resonator via Pound-Drever-Hall locks of 1~MHz loop bandwidth.  Another feedback loop stabilizes the resonator frequency to an atomic transition in ${}^{85}\mathrm{Rb}$.

The atoms are loaded into the standing-wave optical trap from a microchip-based magnetic trap described elsewhere \cite{Teper06}.  After polarization gradient cooling in the linearly polarized optical trap, we apply a 5.6 G magnetic field along the resonator axis and a circular polarization fraction of 0.5(1) to the trap light.  This combination yields a first-order cancellation of the vector and scalar light shifts, minimizing inhomogeneous broadening of the clock transition. Table \ref{tab:trap} summarizes the characteristics of the atomic cloud in the optical dipole trap.

\begin{table}
  \begin{center}

    \begin{tabular}{|lc|c|c|}
      \hline
Parameter & & $\lambda=780$ nm & $\lambda=851$ nm \\ \hline\hline
Mirror separation & $L$ & \multicolumn{2}{c|}{26.62(1) mm} \\\hline
Mirror curvature radius & $R$ & \multicolumn{2}{c|}{25.04(2) mm} \\\hline
Free spectral range & $\omega_\mathrm{FSR}/(2\pi)$ & \multicolumn{2}{c|}{5632.0(2) MHz} \\\hline
Transverse mode spacing & $\omega_\mathrm{t}/(2\pi)$& \multicolumn{2}{c|}{226.3(3) MHz} \\\hline
Linewidth & $\kappa_\lambda/(2\pi)$ & 1.01(3) MHz & 135(2) kHz \\ \hline
Finesse & $\F_\lambda$ & $5.6(2)\times 10^3$ & $4.2(1)\times 10^4$ \\ \hline
Mode waist & $w_\lambda$ & $56.9(4) \hbox{\textmu}$m & $59.5(5) \hbox{\textmu}$m \\ \hline
Antinode cooperativity & $\eta_{0,\lambda}$ & 0.203(7) & 1.65(4) \\ \hline
    \end{tabular}
    \caption{Resonator parameters.  The mode waists are calculated at the position of the atoms.  Outside this table, all resonator values refer to the probe wavelength $\lambda=780$~nm.}
    \label{tab:resonator}
  \end{center}
\end{table}

\begin{table*}
  \begin{center}

    \begin{tabular}{|l|c|c||l|c|c|}
      \hline
\multicolumn{3}{|l||}{Optical dipole trap} & \multicolumn{3}{l|}{Atomic cloud}\\ \hline\hline
Axial frequency & $\omega_\mathrm{ax}/(2\pi)$  & 550 kHz & Length & $l$ & $1$ mm ($\sim2000$ wells)  \\ \hline
Radial frequency &  $\omega_\mathrm{r}/(2\pi)$ & 1.8 kHz & RMS radius & $\sigma_r$ &  8.1(8) $\hbox{\textmu}$m\\ \hline
Trap Depth & $U_0/h$ & 24(1)~MHz & Radial temperature & $k_B T_\mathrm{r}/h$  & 1.3(2) MHz \\ \hline
    \end{tabular}
    \caption{Characteristics of standing-wave dipole trap and atom cloud. The trap depth and trap frequencies are determined from the intracavity power and mode geometry.  The radial temperature is measured by suddenly releasing the atoms and observing their ballistic radial expansion as a decrease in coupling to the resonator.}
    \label{tab:trap}
  \end{center}
\end{table*}

\section{Detection Setup}\label{sec:locking}

\begin{figure*}
  \includegraphics[width=0.8\textwidth]{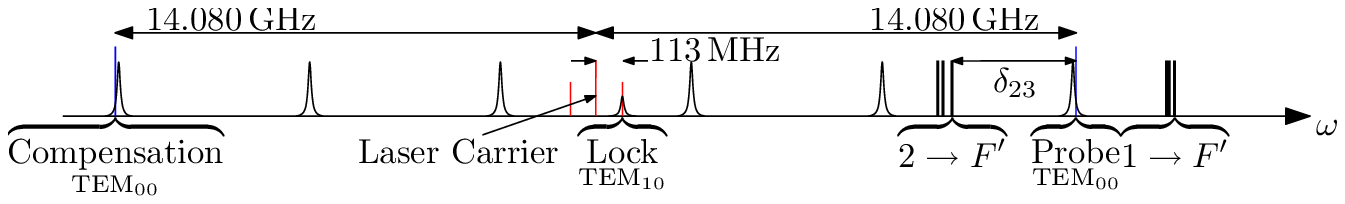}
  \caption{\label{fig:sideband-scheme}Laser stabilization and detection
  scheme, indicating frequencies of carrier and lock sidebands (red) and probe
  and compensation sidebands (blue) relative to cavity resonances and atomic
  transitions.  Not to scale.}
\end{figure*}

We probe the  atoms' index of refraction on the $\mathrm{D_2}$ transition with linear polarization through the optical cavity.  The probe laser carrier lies halfway between two TEM$_{00}$ modes of the resonator.  A broadband electro-optic modulator (model PM-0K5-10-PFA-PFA-780-UL from EOSPACE) is used to generate sidebands for locking and probing (see Fig. \ref{fig:sideband-scheme}).  A lock sideband at 113~MHz, resonant with a TEM$_{10}$ mode, produces the Pound-Drever-Hall error signal.

The probe sideband at $(5\omega_\mathrm{FSR} + \kappa)/2 \approx
2\pi\times14080~\mathrm{MHz}$ lies on the slope of a TEM$_{00}$ resonance with
a detuning of $+2\pi\times 3.57(1)~\mathrm{GHz}$ relative to the atomic $F=2\rightarrow
F'=3$ transition.  The far off-resonant symmetric (compensation) sideband at $-(5 \omega_\mathrm{FSR} +\kappa)/2$ lies on the opposite slope of another TEM$_{00}$ mode, such that the total transmission in the two modes is (ideally) sensitive only to atom-induced shifts of the cavity resonance, but not to frequency jitter of the laser relative to the cavity.

The transmitted power in the TEM$_{00}$ mode is coupled into a
single-mode fiber to filter out the lock sideband and
subsequently detected with overall quantum efficiency $Q_e= 0.43(4)$
on a Si avalanche photodiode (Hamamatsu model S3884). At a typical power of 2~nW for our $T=50~\hbox{\textmu}\mathrm{s}$ long probe pulses, the total photodetection noise (including excess noise of the avalanche photodiode operated at gain $M=13$ \cite{Hamamatsu04}) is a factor of 1.9 in variance above the photocurrent shot noise.

\section{State Preparation}\label{sec:stateprep}
Each cycle of the experiment includes three CSS preparations with the same loaded atoms.  The first CSS preparation precedes the measurements $M_1$ and $M_2$ used to prepare and read out a squeezed state.  The measurement $M_1$ is also used to quantify the unconditional variance of $S_z$ via $y_1=4\Var{M_1}$.  The two subsequent CSS preparations precede the measurements $\tM_1$ and $\tM_2$ used for independent verification of the state preparation noise via $y_2=2\Var{\tM_1-\tM_2}$.

\textbf{First CSS preparation:}  The atoms are optically pumped into $\ket{1,0}$ using $\sigma_+/\sigma_-$-polarized light on the $F=1\rightarrow F'=0$ transition while repumping on $F=2\rightarrow F'=2$. To improve the state purity, a (composite SCROFULOUS \cite{Cummins03}) microwave $\tilde{\pi}$ pulse is subsequently applied on the $\ket{1,0}\rightarrow\ket{2,0}$ transition, all $F=1$ states are emptied using resonant light on the $F=1\rightarrow F'=1$ transition, a second $\tilde{\pi}$ pulse returns atoms from $\ket{2,0}$ to $\ket{1,0}$, and all atoms remaining in $F=2$ ($\sim 12\%$ of the initial atom number) are expelled from the trap using resonant light on the $F=2\rightarrow F'=3$ transition.  After this procedure, more than $99\%$ of the remaining atoms are in the state $\ket{1,0}$.  A $\pi/2$ pulse prepares the atoms in a CSS with $\aver{S_z}=0$, on which the squeezing and readout measurements are performed.

\textbf{Second and third CSS preparations:}  We then proceed, using the same loaded atoms, to prepare a CSS in the $xy$-plane two more times to confirm the CSS projection noise.  In order to compare two identically-prepared CSSs with the same total atom number, we forego the state purification procedure described above, since it leads to a $\sim 12\%$ loss of atoms.  We thereby allow our imperfect optical pumping to leave 12(2) \% of the atoms in $\ket{1,\pm 1}$.  These residual atoms do not contribute to our measurement of $S_z$ because our spin echo technique (see Sec. \ref{sec:spinecho}) cancels any contribution from atoms not addressed by microwaves resonant with the $\ket{1,0}\rightarrow\ket{2,0}$ transition.  Therefore, in Fig.~2, the atom number $N_0$ for the data derived from this pair of preparations (open circles) includes only atoms in $\ket{1,0}$ and $\ket{2,0}$ and is systematically 12$\%$ lower than the atom number used to measure $\Var{M_1}$ and $\Measvar$ (open triangles and solid diamonds).

\subsubsection*{State Preparation Noise}
Figure~2 indicates the presence of technical noise in the state preparation.  The technical noise evident in $\Var{M_1}$ is probably due to slow drifts in microwave power.  An alternative analysis, in which we compare each measurement $M_1$ with the value $M_1^\mathrm{prec}$ in the preceding experiment cycle, yields a fit $2\Var{M_1-M_1^\mathrm{prec}} = 2650(400) + 0.95(23) N_0 + 1(9)\times 10^{-6} N_0^2$, i.e. a result consistent with no contribution from the quadratic term.  We also show in Fig.~2 the variance $y_2=2\Var{\tM_2-\tM_1}$, which is immune to slow drifts in microwave power.  However, after completing this work, we discovered that the state preparation preceding the measurement $\tM_2$ was compromised by an effect of leakage light during that preparation, to which we attribute the small technical noise observed in $y_2$.

\section{Atom-Light Interaction in an Optical Resonator}\label{sec:atomlight}
We summarize the theory of the interaction of a two-level atom with an optical resonator mode at large detuning $\delta\gg\Gamma$, relative to the excited-state linewidth $\Gamma$, from the atomic transition.  The extension to our real system of many atoms with nontrivial level structure follows in Sec. \ref{sec:atomnumber}.

\subsection{Atom-Resonator Coupling, Cooperativity, and Optical Depth}\label{sec:coop}
The atom-resonator coupling $g(\vc{r})=|\vc{d}_{eg}\cdot\vc{E(\vc{r})}|/\hbar$ for an atom at position $\vc{r}=(\rho,z)$ in the Gaussian mode is given by
\begin{equation}
g(\vc{r})^2 = d_{eg}^2 \frac{2\omega_{eg}}{\epsilon_0\hbar\pi w^2 L}e^{-2\rho^2/w(z)^2}\sin^2(k z),
\end{equation}
where $\vc{d}_{eg}$ is the dipole matrix element between the two states $\ket{g}$ and $\ket{e}$, $\omega_{eg}$ is the energy of the transition, $w(z)$ is the mode waist at the position of the atom, and $L$ is the resonator length.  (2$g$ is the vacuum Rabi frequency.)  The coupling $g(\vc{r})$ is related to the atomic excited-state linewidth $\Gamma=\omega_{eg}^3 d_{eg}^2/(3\pi\epsilon_0\hbar c^3)$ and resonator linewidth $\kappa$ by the single-atom cooperativity $\eta(\vc{r})$, the ratio of the scattering rate into the resonator mode to the free-space scattering rate \cite{Vuletic01}:
\begin{equation}\label{eq:eta}
\eta(\vc{r}) = \frac{4 g(\vc{r})^2}{\kappa\Gamma} = \frac{24\F}{\pi k^2 w^2} e^{-2\rho^2/w^2}\sin^2(k z),
\end{equation}
where $\F=\pi c/(L\kappa)$ is the finesse of the resonator and $k=\omega_{eg}/c$ is the probe wavenumber.

The cooperativity is closely related to the resonant optical depth, which for a single atom with scattering cross section $\sigma_\mathrm{sc}$ in a uniform beam of area $A$ in free space is given by $\sigma_\mathrm{sc}/A$.  A light pulse resonant with the cavity passes through the atomic sample on average $2\F/\pi$ times.  For an atom at an antinode of the standing-wave mode, the resonator then enhances the resonant optical depth by a factor of $4\F/\pi$ relative to its value $12/(k^2 w^2)$ on the axis of a running-wave Gaussian beam of waist $w$ in free space, so that $2\eta$ represents the resonator-enhanced single-atom optical depth.

\subsection{Resonator Mode Shift and Back-Action Phase Shift}

We now consider a resonator containing $n$ photons and a single atom in state $\ket{g}$.  The shift $\omega_1 = g^2/\delta = \eta\Gamma\kappa/(4\delta)$ of the resonator mode due to the interaction with the atom is accompanied by an AC Stark shift $n\omega_1$ of the atomic level $\ket{g}$ due to the light; the symmetry between these two effects is readily understood in the dressed-atom picture \cite{Haroche06}.  Since photons are transmitted through the resonator (leaving the resonator in the forward direction) at a rate  $n\kappa/2$, the phase shift of the atomic state $\ket{g}$ per transmitted photon is $2\omega_1/\kappa$.

\section{Population Measurement}
\label{sec:atomnumber}

\subsection{Mode Shift and Effective Atom Number}

In Sec. \ref{sec:atomlight}, we expressed the shift $\omega_1=g^2/\delta$ of a resonator mode coupled to a two-level atom in terms of the cooperativity $\eta$.  In any real atom, at finite detuning, the coupling $g$ is polarization-dependent and must be summed over various excited states.  In terms of the cooperativity $\eta$ of a two-level atom (i.e. the cooperativity on a cycling transition), given by the right-hand side of Eq. \ref{eq:eta}, a single $\Rb$ atom at position $\vc{r}$ occupying state $\ket{F}$ shifts the mode frequency for linearly polarized light on the $\mathrm{D}_2$ transition by an amount
\begin{equation}
\label{eq:shiftF}
\omega_1^{(F)}(\vc{r}) = f\eta(\vc{r})\frac{\Gamma\kappa}{4\delta_F}.
\end{equation}
Here, $\Gamma$ is the excited-state linewidth; $\delta_F$ is an effective detuning from the $\ket{5^2 S_{1/2}, F}\rightarrow\ket{5^2 P_{3/2}, F'}$ transitions averaged over excited hyperfine states $F'$; and $f=\frac{2}{3}$ is the oscillator strength of the $D_2$ line.  For our extended sample of $N_\mathrm{a}$ atoms we define the effective cooperativity $\eta_\mathrm{eff}=f\eaver{\eta^2}/\eaver{\eta}$ and the effective atom number $N_0= N_\mathrm{a} \eaver{\eta}^2/\eaver{\eta^2}$.  Here, $\eaver{}$ denotes an average over the atomic ensemble.  This definition, which yields $N_0 \approx \frac{2}{3}N_\mathrm{a}$, is chosen such that the projection noise variance satisfies the usual condition for a uniform sample,
\begin{equation}
\frac{\var{N_0}}{N_0} = 1.
\end{equation}
The mode shift due to an ensemble in state $\ket{F}$ is $\omega^{(F)}\nolinebreak =\nolinebreak N_0 \eta_\mathrm{eff}\Gamma\kappa/(4\delta_F)$, where $\eta_\mathrm{eff}$ is related to the cooperativity $\eta_0=24\F/(\pi k^2 w^2)$ of a maximally coupled atom by
\begin{equation}
\label{eq:etaeff}
\frac{\eta_\mathrm{eff}}{\eta_0} = f\frac{\eaver{\sin^4kz}}{ \eaver{\sin^2kz}} \frac{w^2+4\sigma_r^2}{w^2+8\sigma_r^2} =  0.47(1)
\end{equation}
for our cloud of radius $\sigma_r \ll w$.

\subsection{Measurement of $S_z$}

\begin{table*}
  \begin{center}
    \begin{tabular}{|l|c|c|c|}
      \hline
      Resonance & $\delta_{23}/(2\pi~\mathrm{GHz})$ &
      \multicolumn{2}{|c|}{$\mbox{Cavity Shift per Effective Atom}
      /(\kappa\times 10^{-6})$} \\ \cline{3-4}
      & & \ket{F=1,m_F=0} & \ket{F=2,m_F=0} \\ \hline\hline
      Probe & 3.57(1) &  -49(2) & +39(1) \\
      Lock & -10.40(1) & -0.6(1) & -1.0(2) \\
      Compensation & -24.59(1) & -4.6(2) & -5.8(2) \\ \hline
    \end{tabular}
    \caption{Frequency shifts per effective atom in either of the clock states for relevant resonator modes.  $\delta_{23}$ is the detuning of the given mode from the
    $\ket{5^2S_{1/2},F=2}\rightarrow \ket{5^2P_{3/2},F'=3}$ atomic transition. The calculated shifts include the excited-state hyperfine structure, as well as the spatial overlap of the cloud with the mode.}
    \label{tab:shifts}
  \end{center}
\end{table*}

The probe sideband is tuned to the frequency between the $|5^2 S_\mathrm{1/2}, F=1\rangle\rightarrow|5^2 P_\mathrm{3/2}\rangle$ and $|5^2
S_\mathrm{1/2}, F=2\rangle\rightarrow|5^2 P_\mathrm{3/2}\rangle$ transitions at which the atom-induced differential frequency shift between the probe and compensation sidebands,
\begin{equation}
\label{eq:shift}
\omega = \frac{(N_2-N_1)\eta_\mathrm{eff}\Gamma\kappa}{4\delta'},
\end{equation}
is proportional to the effective-atom population difference $N=N_2-N_1$ between the hyperfine states $F=1,2$, but independent of the total atom number $N_1+N_2$. Here $\delta' = 2\pi\times 3200(10)~\mathrm{MHz}$; see Table \ref{tab:shifts} for details. For our cloud geometry and probe polarization, the differential mode shift per effective atom is $d\omega/dN=2\pi\times 45(1)~\mathrm{Hz/atom}=4.5(2)\times 10^{-5}~\kappa/\mathrm{atom}$. The mode shifts due to projection noise on $N$ are much smaller than $\kappa/2$, leaving the resonator transmission in the linear regime, such that the change in transmitted power is directly proportional to $S_z=N/2$.  Nevertheless, in all measurements of $S_z$ we take into account the full Lorentzian lineshape of the resonator transmission, allowing the same procedure to be used for measuring projection noise, contrast, and total atom number. The cavity linewidth $\kappa/(2 \pi)=1.01(3)$~MHz at the probe wavelength is accurately measured by tuning the probe sideband over the TEM$_{00}$ resonance of the empty cavity and measuring probe transmission.

\subsection{Measurement of $N_0$}
\begin{figure}
\includegraphics[width=\columnwidth]{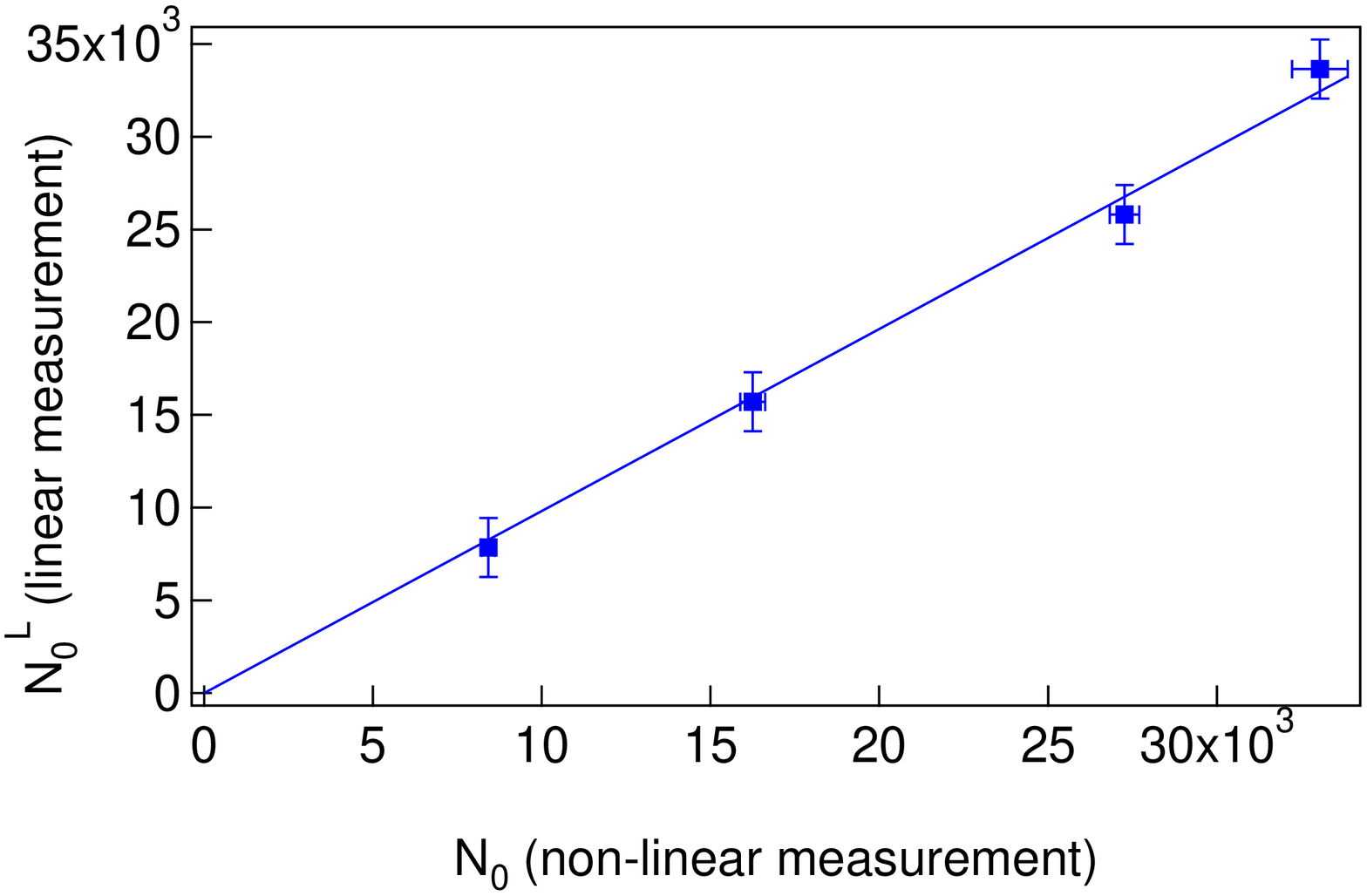}
\caption{Comparison of two methods of determining effective atom number, as described in the text.  The vertical error bars indicate the uncertainty in positioning the probe sideband on cavity resonance.  The horizontal error bars arise because the atom number $N_0$ is the average over a different set of loading cycles from those in which $N_0^L$ is measured.}
\label{fig:ShiftVsShift}
\end{figure}

At the end of each experiment cycle, we measure the effective atom number $N_0$.  We determine the atom number by pumping all atoms first into $F=2$, then into $F=1$, and in each case measuring the resonator transmission.  Although the resonator mode shifts $\omega^{(F)}$ are linear in $N_0$, they are on the order of $\kappa$ for our typical atom numbers, so that the transmitted power is a non-linear function of $\omega^{(F)}$ which we must invert to obtain the atom number.  To verify our determination of the resonator shift from the non-linear transmission signal, we additionally measure the average mode shift over several cycles of the experiment by finding the probe sideband frequencies $\omega_p^{(F)}$ that maximize the probe transmission when all atoms are pumped into hyperfine state $F$.  By thus directly measuring the mode frequency shift, we obtain a linear measure of effective atom number $N_0^L = (\omega_p^{(2)}-\omega_p^{(1)})/2/(2 \pi \times 45~ \mathrm{Hz})$.  In Figure \ref{fig:ShiftVsShift} we plot $N_0^L$ against the average atom number $N_0$ extracted from the nonlinear transmission signal.  The fit $N_0^L = 0.98(4) N_0$ with reduced $\chi^2=0.3$ indicates that the two measurements are in good agreement.

At the large atom number $N_0 = 3.3(2)\times 10^4$ where we calculate squeezing parameters, the dominant uncertainty in $N_0$ arises from sensitivity to the initial placement of the probe and compensation sidebands.  In extracting $N_0$ from the transmission, we assume that we have correctly placed the probe and compensation sidebands at $\pm\kappa/2$ detuning from cavity resonance when $S_z=0$.  We monitor the placement of the sidebands in each cycle of the experiment by shifting the frequency of the laser relative to the resonator by $+\kappa$ and $-\kappa$ from the usual configuration and measuring the transmission in each case.  Hence, we are confident that any systematic error in the placement of these sidebands is less than the shot-to-shot fluctuations.  We therefore always estimate the uncertainty in $N_0$ by the standard deviation of the calculated $N_0$ values.

\subsection{Experimental Verification of Atom-Resonator Interaction}
\label{sec:PhotonPhase}

\begin{figure}
\includegraphics[width=\columnwidth]{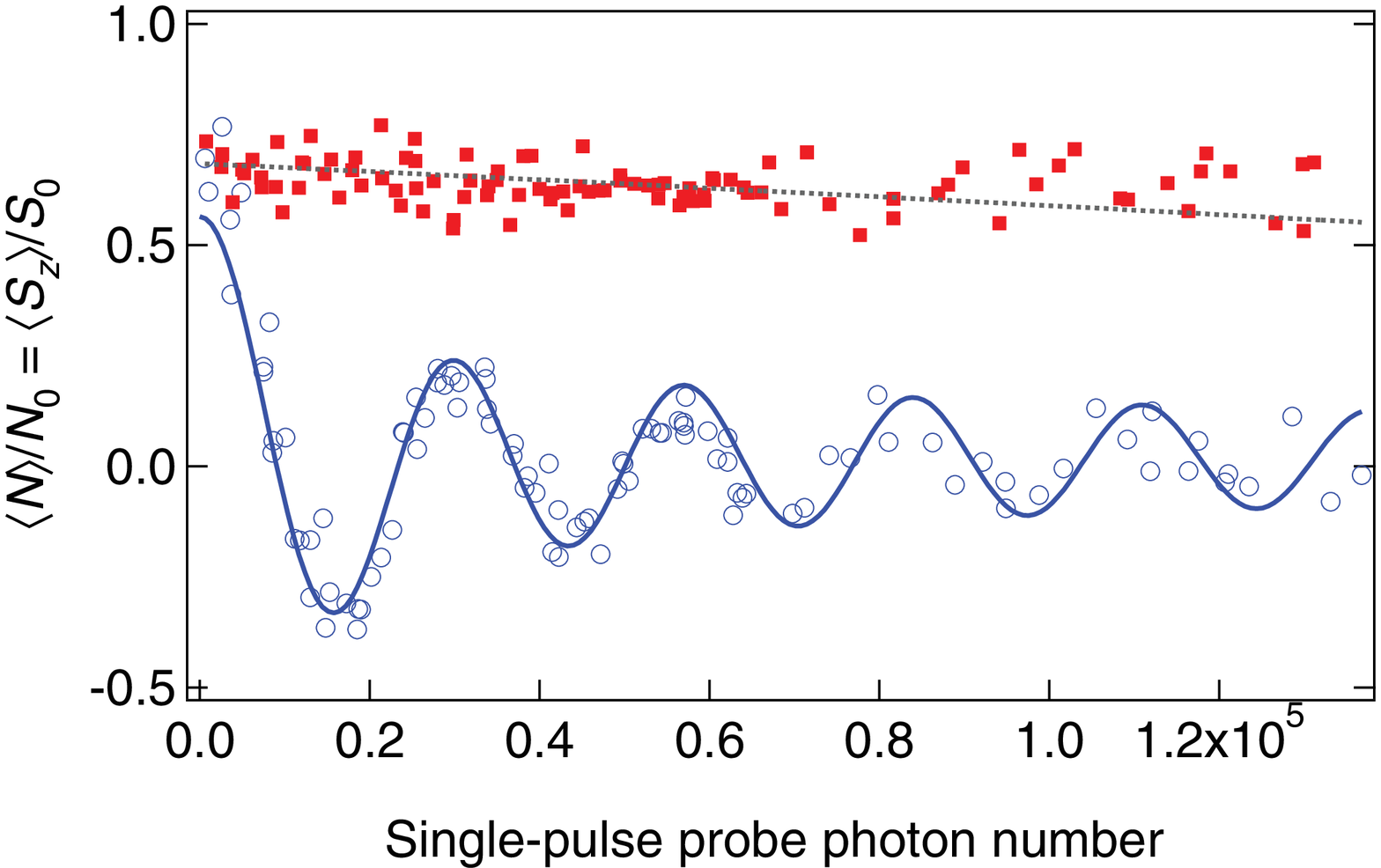}
\caption{Measurement of the atom-photon interaction. A
probe pulse of varying photon number $p$ is inserted into a Ramsey
sequence, resulting in a differential light shift between atomic states. The light-induced phase shift and decoherence (open circles) can be suppressed by a spin echo technique where a microwave $\pi$ pulse is inserted between two probe pulses (solid squares).}
\label{PhotonPhase}
\end{figure}

The calculated mode shift per atom, Eqs. \ref{eq:shiftF} and \ref{eq:shift}, is used to convert measured transmission into atom number.  We verify it by measuring the complementary atomic phase shift $\phi=2(\omega_1^{(2)}-\omega_1^{(1)})/\kappa$ induced between states $\ket{1}$ and $\ket{2}$ by a single probe photon transmitted through the resonator.  We determine the phase shift $\phi_0$ of a maximally-coupled atom by means of a Ramsey measurement \cite{Ramsey50}, applying an optical probe pulse of variable duration between two microwave $\pi/2$ pulses.  The population difference $2S_z$, measured via the resonator shift (Fig.~\ref{PhotonPhase}), is an oscillatory function of the transmitted probe photon number $p$. The oscillation is damped due to inhomogeneous light shifts.  For an ensemble of atoms on the resonator axis evenly distributed with respect to the probe standing wave, a spin state prepared along the $x$-axis of the Bloch sphere acquires, after the interaction,
\begin{align}
  \eaver{S_x} \propto&\frac{\int_0^{2\pi}\cos\left(p\,\phi_0\sin^2(kz)\right)\sin^2(kz)\,dz}
  {\int_0^{2\pi}\sin^2(kz)\,dz}\nonumber\\
  =& J_0\left(u\right)\cos\left(u\right)-J_1\left(u\right)\sin\left(u\right),
\end{align}
where the $J_n$ are Bessel functions of the first kind and $u=p\phi_0/2$.  From a fit of this form we extract $\phi_0 = 230(20)~\hbox{\textmu}\mathrm{rad}$. (The phase shifts due to lock and compensation light are negligible.)   A fit to a full numerical model including the radial cloud size yields $\phi_0^\mathrm{meas} = 250(20)~\hbox{\textmu}\mathrm{rad}$, in excellent agreement with the value $\phi_0^\mathrm{calc}= 253(8)~\hbox{\textmu}\mathrm{rad}$ calculated from cavity parameters.

\section{Spin Echo Sequence}
\label{sec:spinecho}

We use a spin echo technique to reduce the probe-induced inhomogeneous broadening (Fig. \ref{PhotonPhase}). All probe light is applied in two $50~\hbox{\textmu}\mathrm{s}$ long pulses separated by a composite $\tilde{\pi}$ pulse, consisting of a sequence $R_{\pi/3}(\pi)R_{-\pi/3}(\pi)R_{\pi/3}(\pi)$ of three simple microwave $\pi$ pulses, where the subscripts indicate phases
chosen to compensate variations in pulse area \cite{Vandersypen04}. The spin echo is optimized at a probe pulse separation of $330(20)~\hbox{\textmu}\mathrm{s}$, corresponding to a half-period of the radial trap oscillation.

\section{Data Analysis} \label{sec:snred_c}

\subsection{Noise Model}\label{sec:noisemodel}

\begin{figure}
\includegraphics[width=\columnwidth]{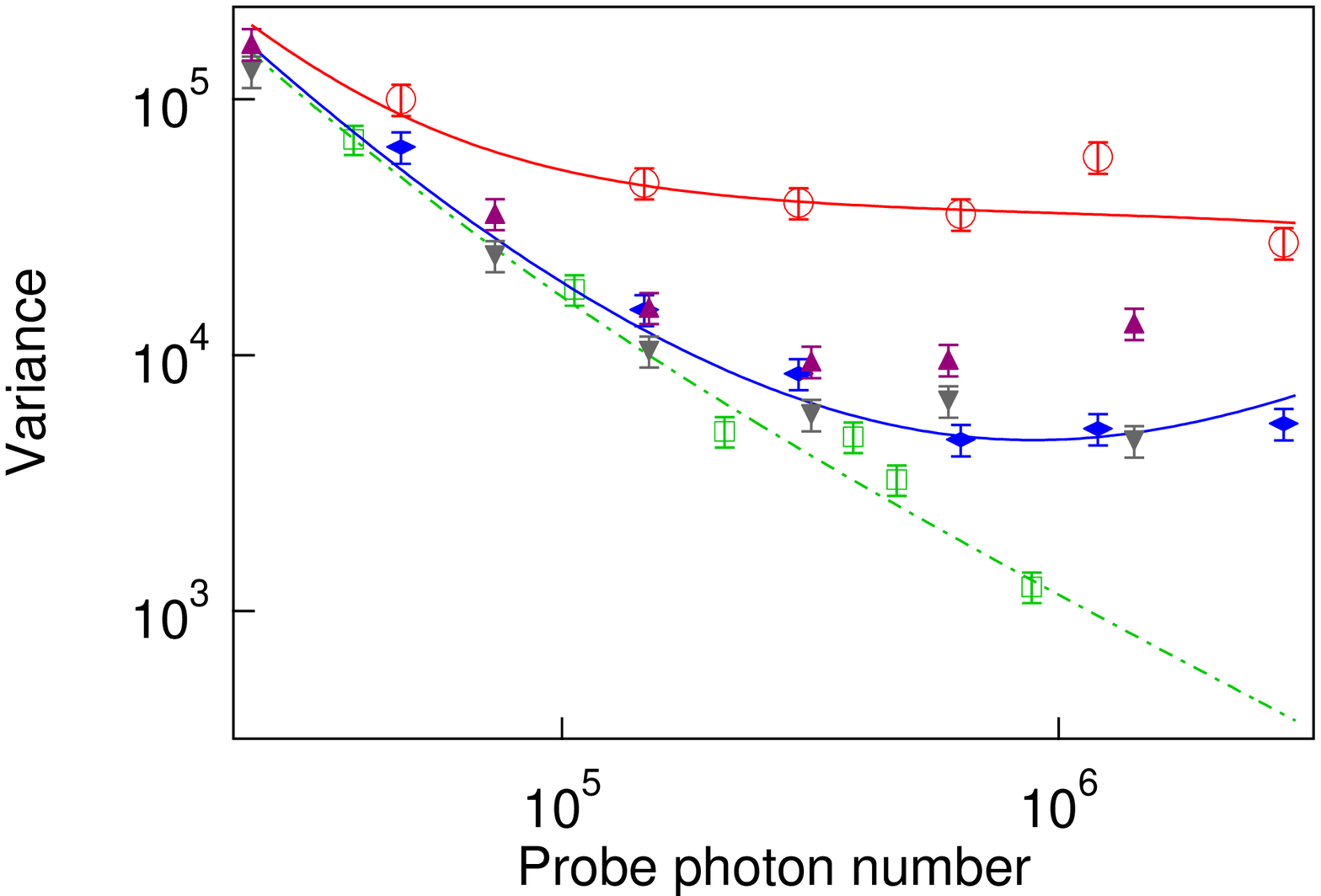}
\caption{Comparison of observed variances with noise model.  Shown as a function of probe photon number $p$ are the measurement variance in atom number units, $4\Measvar=2\Var{M_1-M_2}$, with (solid blue diamonds) and without (open green squares) atoms; and the variance $4\Var{M_1}$ with atoms, which includes projection noise.  In addition, variances $2\Var{M_{1+}-M_{2+}}$ (solid gray inverted triangles) and $2\Var{M_{1-}-M_{2-}}$ (solid purple triangles) using a single probe pulse from each measurement (as shown in Fig. \ref{fig:probe_pulses}) are plotted as a function of single-pulse probe photon number $p/2$.  Curves correspond to the noise model described in Sec. \ref{sec:noisemodel}.}
\label{fig:var_vs_p}
\end{figure}

The spin measurement variance $\Measvar=\Var{M_1-M_2}/2$ as a function of probe photon number $p$ used in the measurement (Fig. \ref{fig:var_vs_p}) is well described by assuming the following independent noise contributions to $4\Measvar = b_{-2} p^{-2}+b_{-1} p^{-1}+b_0 p^0 +b_1 p^1$: electronic noise of the detector scaling as $\Measvar\propto p^{-2}$; photon shot noise and avalanche excess noise scaling as $\Measvar\propto p^{-1}$; technical noise that is independent of photon number, including the effects of imperfect microwave rotations in the spin echo procedure; and noise due to photon (Raman) scattering, $\Measvar\propto p$.  We quantify each of the known noise contributions:

\enlargethispage{\baselineskip}
\begin{itemize}
\item{\textbf{Photon shot noise and avalanche excess noise:} We calculate the photocurrent noise due to the probe and compensation light, taking into account both the shot noise of the light detected with a quantum efficiency $Q_e= 0.43(4)$ and the excess noise factor $f_\mathrm{APD}=1.9(4)$ of the avalanche photodiode operated at a gain of 13 \cite{Hamamatsu04}, obtaining $b_{-1} = 2(f_\mathrm{APD}/Q_e)(dN/d(2\omega/\kappa))^2=1.1(3)\times 10^{-9}$}.
\item{\textbf{Electronic noise:} From a fit to the noise measured in the absence of atoms (open green squares in Fig.~\ref{fig:var_vs_p}) in which we constrain the coefficient $b_{-1}$ to the value calculated above, we obtain an electronic noise contribution $b_{-2}=6(1)\times 10^{13}$, most of which is attributable to the Johnson noise of the transimpedance gain resistor in the photodetection circuit.}
\item{\textbf{Microwave infidelity:} The $\tilde{\pi}$ pulse used in the spin echo produces at least $98(1)\%$ inversion.  We model the imperfect $\tilde{\pi}$ pulse as a perfect one combined with an incoherent process that flips on average $\mu=2(1)\%$ of the spins, yielding $b_{0,\mu} = \mu N_0$.  We treat the errors as incoherent because the atomic phase is inhomogeneously broadened by $(\phi_0/2)p = 1.3\times 10^{-4} p$ radians when the microwaves are applied.  At the optimum photon number $p=3\times 10^5$ for squeezing, the Ramsey contrast remaining after a single probe pulse is only $10(3)\%$ (see Fig. \ref{PhotonPhase}).  We briefly address possible coherent noise processes below (Sec. \ref{sec:cohbound}), placing an upper bound on the effect of such processes.}
\item{\textbf{Raman scattering:} In our system, the probability of a Raman scattering event is $\PRam=5.6\times 10^{-8}$ per probe photon transmitted through the resonator.  This value, calculated including the full excited-state and ground-state hyperfine structures, includes probabilities $\PRam=\flipin+\shift+\flipout$ corresponding to three types of scattering events: those which change $F$ but not $m_F$, those which change $m_F$ but not $F$, and those which change both $F$ and $m_F$, respectively.  To first order in these probabilities, the total contribution of Raman scattering to the measurement variance is $b_1=(4/3\flipin+1/2\shift+1/3\flipout)N_0=4.7\times 10^{-8} N_0$ per probe photon.  Section \ref{sec:scattderiv} outlines the derivation of this expression for $b_1$.}
\end{itemize}

Figure \ref{fig:var_vs_p} shows a fit of the above model (solid blue curve) to the observed measurement variance in atom number units, $4\Measvar$, at $N_0=3.3(2)\times 10^4$ and variable photon number $p$ (solid blue diamonds).  We fix the noise contributions enumerated above and leave free a term $b_{0,tech}$ to account for technical noise that is independent of probe photon number but not due to microwave infidelity.  The value $b_{0,tech} = 1400(400) = 0.04(1)N_0$ obtained from the fit may be due to frequency jitter of the laser relative to the cavity that is incompletely canceled by the compensation sideband; or to technical noise in the probe light level, e.g. from fluctuations in the coupling of the light to the cavity.  The data show very good agreement with our noise model with this single free parameter $b_{0,tech}$, all other parameters being independently measured or calculated as described above.

\begin{figure}
\includegraphics[width=\columnwidth]{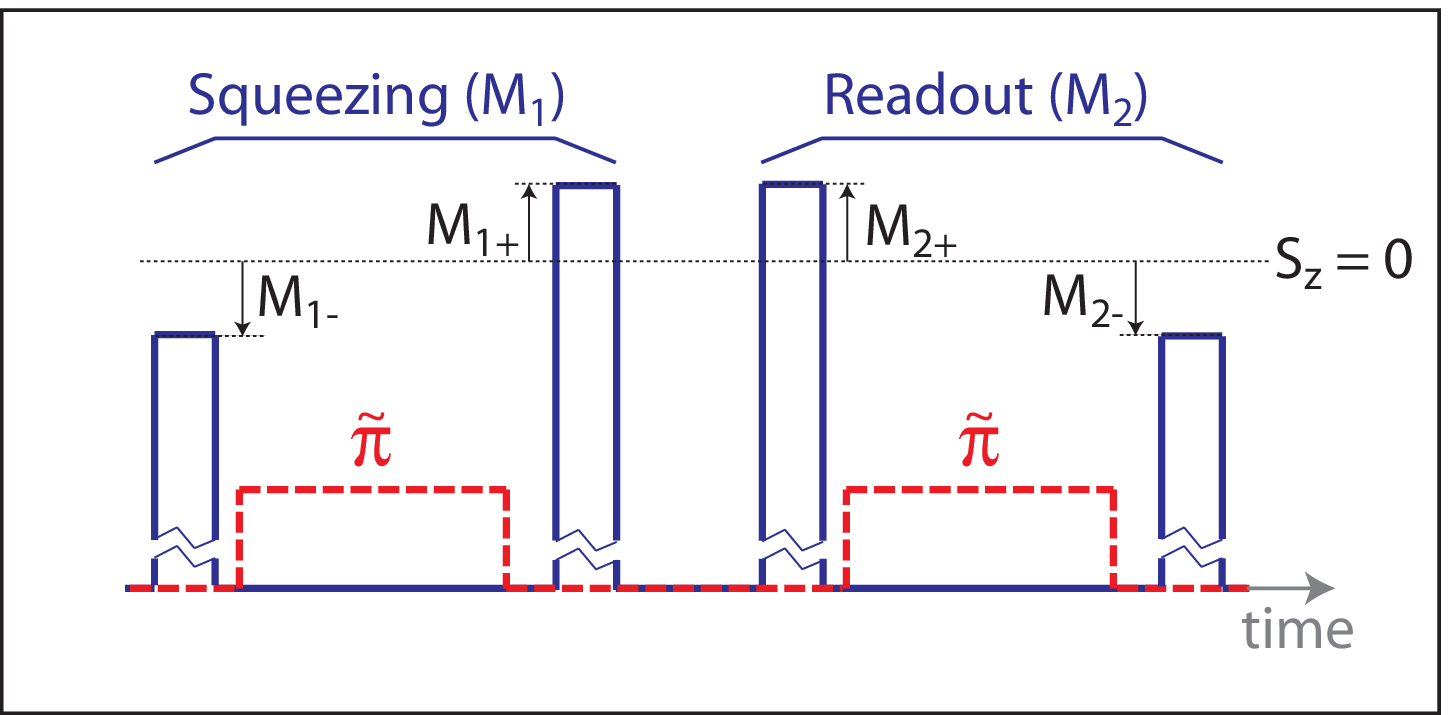}
\caption{Measurements $M_1$ and $M_2$ of $S_z$ for spin squeezing and readout.  Each measurement is obtained from the probe light transmitted (solid blue line) in two pulses, before and after applying microwaves (dashed red line) to perform a $\tilde{\pi}$ rotation.  Specifically, each measurement can be expressed as the average $M_i=(M_{i+}+M_{i-})/2$ of two single-pulse measurements $M_{i\pm}=\pm\omega_{i\pm}/|2d\omega/dN|$, where $\omega_{i\pm}$ are the atom-induced shifts of the cavity resonance deduced from the transmitted probe light.  With each $\tilde{\pi}$ rotation, we switch the sign convention for $S_z=\pm(N_2-N_1)/2$ to compensate for the population exchange between states $\ket{1}$ and $\ket{2}$.}
\label{fig:probe_pulses}
\end{figure}

The distinction between the two contributions to $b_0 = b_{0,\mathrm{tech}}+b_{0,\mu}$ is verified by also plotting measurement variances (in atom number units) $2\Var{M_{1-}-M_{2-}}$ and $2\Var{M_{1+}-M_{2+}}$ obtained by comparing either the first and last or the second and third of the four probe pulses $M_{i\pm}$ constituting the two measurements $M_i$, $i\in \{1,2\}$; see Fig. \ref{fig:probe_pulses}.  Whereas two $\tilde{\pi}$ pulses are applied between $M_{1-}$ and $M_{2-}$, no microwaves are applied between $M_{1+}$ and $M_{2+}$.  Thus, fits (not pictured) to $2\Var{M_{1-}-M_{2-}}$ and $2\Var{M_{1+}-M_{2+}}$ reveal the microwave infidelity $\mu_\mathrm{fit}=(b_0^{-}-b_0^{+})/(4 N_0)=0.03(1)$, consistent with the independently determined value $\mu=0.02(1)$.  (The value $\mu_\mathrm{fit}$ also includes a small contribution $<0.01$ due to photons scattered by the lock light.)

In addition, we plot $4\Var{M_1}$ (open red circles in Fig. \ref{fig:var_vs_p}) and a curve given by the following expression: $4\Var{M_1} = b_{-2}/p^2+b_{-1}/p+b_{0,\mathrm{tech/prep}}+(1-\mu-[2\flipin/3+\shift/2+2\flipout/3]p)N_0$ (see Sec. \ref{sec:scattderiv}).  Here, we fit the term $b_{0,\mathrm{tech/prep}}=0.14(7)N_0$ to allow for technical noise in the state preparation but constrain all other parameters to the values given above.

The red curve in Fig.~3 is derived by combining the fits to $4\Measvar$ and $4\Var{M_1}$ described above with a fit to the contrast, described in Sec. \ref{sec:contrast}, in accordance with Eq. \ref{eq:zetaW} below.  The dashed blue curve in Fig.~3 corresponds to the same expression without the contrast factors.


\subsubsection{Bound on Coherent Microwave Errors}
\label{sec:cohbound}

We have noted above that any noise process associated with coherent microwave pulse errors in the spin echo is highly suppressed by inhomogeneous broadening of the atomic phase before the spin-echo-induced rephasing.  For example, the drifts of $0.4 \%$ in microwave power ($0.2 \%$ in Rabi frequency) which we infer from our state preparation noise $\Prepvar$ would lead to rotation errors substantially less than $\delta\phi_\mathrm{max}=2\times 10^{-3}\pi$ in our composite $\tilde{\pi}$ pulse, which is designed to compensate for microwave power errors.  At the optimum squeezing photon number $p=3\times 10^5$, where the interference contrast is $\mathcal{C}_\mathrm{SE}\approx 10 \%$ when the spin-echo microwaves are applied (see Fig. \ref{PhotonPhase}), the resulting normalized variance in $S_z$ would be at most $\var{S_z}_\mathrm{coh}/\CSSvar < \delta\phi_\mathrm{max}^2\mathcal{C}_\mathrm{SE}^2 N_0 = 4\times 10^{-7} N_0\approx 0.01$ at $N_0=33000$.  This is much smaller than the normalized spin noise $\Condvar/\CSSvar = 0.20(3)$ at the same photon number.

Note further that, while the calculation of conditional spin noise from the spin measurement variance $\Measvar = \Var{M_1-M_2}/2$ obscures measurement errors that are perfectly correlated, it overestimates by a factor of two the effect of measurement errors that are perfectly anticorrelated.  Since correlated microwave errors can yield either correlated or anticorrelated measurement errors depending upon the atomic phase, which in turn depends sensitively on the number of probe photons applied before the $\tilde{\pi}$ pulse in each measurement, we expect a further suppression of any correlated noise below the bound given here.

\subsubsection{Derivation of Spin-Flip Terms}
\label{sec:scattderiv}
This section outlines our derivation of the contributions of incoherent spin flips to the variances $4\Measvar$ and $4\Var{M_1}$ in our noise model above.  Both $4\Measvar=\Var{M_{1+}+M_{1-}-M_{2+}-M_{2-}}/2$ and $4\Var{M_1}=\Var{M_{1+}+M_{1-}}$ are composed of covariances $\Cov{M_{i\alpha}}{M_{j\beta}}$ of single-pulse measurement outcomes, where $i,j\in\{1,2\}$ and $\alpha,\beta\in\{+,-\}$ (see Fig. \ref{fig:probe_pulses}).  To evaluate these covariances, we first express each single-pulse measurement $M_{i\alpha}$ as
\begin{equation}\label{eq:Mintegral}
M_{i\alpha} = \frac{\alpha\int_0^{T} \omega_{i\alpha}(t) \, dt}{2\abs{d\omega/dN}T},
\end{equation}
where $\omega_{i\alpha}(t)$ is the atom-induced cavity shift at a time $t$ from the beginning of pulse $i\alpha$; we here neglect errors in determining the atom-induced cavity shift due to photon shot noise and technical noise, as these are uncorrelated with the atomic state and can be treated separately.  In the absence of Raman scattering, each $\omega_{i\alpha}$ is constant in time; and if the spin-echo $\tilde{\pi}$ pulses have perfect fidelity, then $\omega_{i-}=-\omega_{i+}$.  Raman scattering and microwave infidelity cause deviations from this ideal behavior, so that
\begin{equation}\label{eq:covomega}
\Cov{\omega_{i\alpha}(t)}{\omega_{j\beta}(t')} \approx \alpha\beta\abs{\frac{d\omega}{dN}}^2 N_0\left[1-2r_{i\alpha,j\beta}(t,t')\right].
\end{equation}
Here, for $\alpha=\beta$, we define $r_{i\alpha,j\beta}(t,t')$ as the probability for an atom to be in a different hyperfine state $F$ at time $t'$ of pulse $j\beta$ than at time $t$ of pulse $i\alpha$; for $\alpha=-\beta$, meaning that there is exactly one microwave pulse between $t$ and $t'$, $r_{i\alpha,j\beta}(t,t')$ is the probability for the hyperfine state to be the same at both times.  We have made the approximations that the initial state preparation is projection-noise limited; and that $r_{i\alpha,j\beta}(t,t')\ll 1$, meaning that all the incoherent-spin-flip probabilities---namely, the Raman scattering probabilities $p\flipin$, $p\shift$, and $p\flipout$ and the microwave infidelity $\mu$---are small.  Each expression for $r_{i\alpha,j\beta}(t,t')$ is then, to lowest order, a linear combination of these probabilities with coefficients that depend on how many $\tilde{\pi}$ pulses are applied between pulse $i\alpha$ and pulse $j\beta$.  For example, $r_{1-,1+}(t,t')\approx\frac{t'-t}{2T}p\flipin+\frac{p}{2}\shift+\frac{t+t'}{2T}p\flipout+\mu$.  This expression and similar expressions for the other $r_{i\alpha,j\beta}(t,t')$ are used to evaluate the covariances $\Cov{M_{i\alpha}}{M_{j\beta}}$ using Eqs. \ref{eq:Mintegral} and \ref{eq:covomega}.  We thus obtain the terms $(4/3\flipin+1/2\shift+1/3\flipout+\mu)N_0$ in our model for $4\Measvar$ and the terms $(1-\mu-[2\flipin/3+\shift/2+2\flipout/3]p)N_0$ in our model for $4\Var{M_1}$.

These results can be understood qualitatively as follows.  Incoherent spin flips increase the spin measurement variance $\Measvar=\Var{M_1-M_2}/2$ by reducing the correlation between the squeezing and readout measurements $M_1$ and $M_2$.  However, they diminish the projection noise observed in the variance $\Var{M_1}$ of a single measurement, since any scrambling of spins allows the measurement to average over the different ensemble spin states at different times.

\subsection{Derivation of Metrological Squeezing Parameter}\label{sec:metrologyfactor}

In interpreting $\Measvar$ as a measurement uncertainty, and in deriving the quantum uncertainty $\Condvar$ of the state prepared by the squeezing measurement, we make assumptions of uncorrelated noise in the two measurements $M_1$ and $M_2$.  These assumptions are justified by the noise model in Sec. \ref{sec:noisemodel}.  The dominant noise contributions at the optimum squeezing point---photodetector noise and technical noise attributable to frequency jitter between laser and cavity---are all uncorrelated between the measurements.  (Any frequency shaking that is slow enough to be common to both $M_1$ and $M_2$ is also common to the cavity shifts $\omega_+$ and $\omega_-$ before and after the spin echo and thus does not affect our measurement of $S_z\propto\omega_+-\omega_-$.)  The remaining noise, due to changes in the atomic state via Raman scattering or microwave $\tilde{\pi}$ pulse infidelity, is correlated with the atomic state and hence affects the correlation between the two measurement outcomes.  We here generalize the derivation of the conditional spin noise and metrological squeezing parameter to encompass the small effects of such spin-flip noise, thereby fully accounting for all processes in our noise model.

The goal of our analysis is to evaluate $\zetaW=2\Condvar S_{\mathrm{in}}/(\meanS^2)$ using results of our measurements $M_1$ and $M_2$.  Since the measurements can change the atomic state, we define $\Szf$ as the value of $S_z$ at the end of the first measurement $M_1$.  The conditional quantum uncertainty $\Condvar$ of the state prepared by the first measurement is found by minimizing $\Var{\Szf-wM_1}$ with respect to the weight $w$ given to the measurement information \cite{Leuchs03, Appel09}:
\begin{equation}\label{eq:varSz1}
\Condvar = \mathrm{min}_w\{\Var{\Szf-wM_1}\}.
\end{equation}
The minimum occurs at $w=\Cov{\Szf}{M_1}/\Var{M_1}$ and is given by \cite{Leuchs03}
\begin{equation}\label{eq:varSz2}
\Condvar = \Var{\Szf}\left(1-\frac{\Cov{\Szf}{M_1}^2}{\Var{\Szf}\Var{M_1}}\right).
\end{equation}

As described in Sec. \ref{sec:noisemodel}, the noise in our system is well described by noise that is uncorrelated with $\Szf$ and by incoherent spin flips.  We begin with the simplifying assumption that all Raman scattering events transfer atoms between the two clock states.  In this model, for identical measurements $M_1$ and $M_2$,
\begin{align}\label{eq:cov}
\Cov{\Szf}{M_1}^2 &= \Cov{\Szf}{M_1}\Cov{\Szf}{M_2}\nonumber\\
&= \Cov{M_1}{M_2}\Var{\Szf},
\end{align}
since the only correlations between the two measurements are due to their linear dependence on the common value $\Szf$ of $S_z$ at the end of the first measurement and the beginning of the second measurement.  Here, $\Var{\Szf}$ represents the unconditional variance of an ideal, noiseless readout following the squeezing measurement.

We would like to express $\Var{\Szf}$ in terms of measured quantities.  We define $\Measvar\equiv \Var{M_1-M_2}/2$ and $\Prepvar\equiv\Var{M_1}-\Measvar=\Cov{M_1}{M_2}$, where the latter equality holds because $\Var{M_1}=\Var{M_2}$.  In the limit where the measurement does not change the atomic state and all measurement noise is uncorrelated, $\Prepvar$ represents the unconditional variance of $S_z$, including the CSS projection noise and any technical noise in the initial state preparation.  To allow for a measurement that induces spin flips, we use Eq. \ref{eq:cov} to relate $\Prepvar$ to $\Var{\Szf}$, obtaining
\begin{equation}
\frac{\Prepvar}{\Var{\Szf}} = \left(\frac{\Cov{M_i}{\Szf}}{\Var{\Szf}}\right)^2 \equiv (1-\epsilon_p)^2.
\end{equation}
To lowest order in the spin-flip probabilities associated with Raman scattering and microwave pulse infidelity, $\epsilon_p \approx p\flipin +\mu$.  The measured unconditional variance $\Prepvar$ is lower than the true variance $\Var{\Szf}$ by the factor $(1-\epsilon_p)^2$ because those spins that are flipped partway through the first measurement contribute less to the measurement (of the time-averaged spin) than to the value $\Szf$ at the end of the measurement.  Hence,
\begin{equation}\label{eq:varSz3}
\Condvar =\frac{\Measvar\Prepvar}{(1-\epsilon_p)^2 (\Prepvar+\Measvar)}.
\end{equation}
The factor $(1-\epsilon_p)^2$ has a 0.3~dB effect, smaller than the statistical uncertainty, on the minimum normalized spin noise $\sigma^2\equiv\Condvar/\CSSvar$, at $p=6.4(6)\times 10^5$.  We do not include this factor in Fig.~3, but we correct for it in reporting the minimum normalized spin noise $\sigma^2$, as well as the entanglement parameter $\zetaKU$ at $p=3\times 10^5$.  We show below that the metrological squeezing parameter $\zetaW$ is independent of $\epsilon_p$.

Note that if we define the measurement strength $\kappa_\mathrm{meas}\equiv\Delta{S_z}_\mathrm{prep}/\Delta{S_z}_\mathrm{meas}$, then for a projection-noise-limited state preparation ($\Prepvar=\CSSvar$) and a measurement that does not change the atomic state ($\epsilon_p=0$) Eq. \ref{eq:varSz3} reduces to the expression $\sigma^2=1/(1+\kappa_\mathrm{meas}^2)$ given by Appel \textit{et al.} \cite{Appel09}.  Whereas Appel \textit{et al.} define the measurement strength for a photon-shot-noise-limited measurement, we have here extended the definition to include more general noise.

We wish to determine the metrological squeezing parameter $\zetaW=2\Condvar S_{\mathrm{in}}/(\meanS^2)$ where $\meanS$ is the length of the mean spin vector  at the end of the squeezing measurement.  The Rabi oscillation curve from which we obtain the contrast (see Sec. \ref{sec:contrast}) is measured using our standard readout, identical to the squeezing measurement.  The contrast $\mathcal{C}(p)=\meanS/S_0$ that would be observed in an ideal readout of $\Szf$ is related to our measured contrast $\mathcal{C}_\mathrm{meas}(p)$ by $\mathcal{C}_\mathrm{meas}(p)/\mathcal{C}(p)= \Cov{M_2}{\Szf}/\Var{\Szf}=1-\epsilon_p$.  Hence,
\begin{align}\label{eq:zetaW}
\zetaW &= \frac{\mathcal{C}_{\mathrm{in}}}{\mathcal{C}^2}\frac{\Condvar}{\CSSvar}\nonumber\\
&=\frac{\mathcal{C}_{\mathrm{in}}}{\mathcal{C}_\mathrm{meas}^2}\frac{2\Measvar\Prepvar}{S_0(\Prepvar+\Measvar)}.\end{align}
The determination of $\mathcal{C_\mathrm{in}}\equiv S_\mathrm{in}/S_0$ is discussed in Sec. \ref{sec:contrast}.  Using Eq. \ref{eq:zetaW}, we obtain the metrological squeezing parameter $\zetaW$ directly in terms of measured variances.  Outside this section (\ref{sec:metrologyfactor}), we do not distinguish between $\mathcal{C}$ and $\mathcal{C}_\mathrm{meas}$.

The interesting fact that our evaluation of the metrological squeezing parameter $\zetaW$, proportional to the squared noise-to-signal ratio, is not affected by photon scattering into free space can be understood as follows: A longer and longer measurement of projection noise fluctuations produced by photon scattering into free space during the measurement, as described by $\epsilon_p$, results in a reduction of the measured spin noise. This would lead one to underestimate the spin noise by a factor $(1-\epsilon_p)^2$, Eq. \ref{eq:varSz3}. However, free-space scattering also reduces the observed contrast, such that the signal-to-noise ratio and the metrological squeezing parameter $\zetaW$ do not depend on $\epsilon_p$.

In the above derivation, we have assumed that all Raman scattering events transfer atoms between the two clock states.  Raman scattering out of the clock states introduces a small correction to Eq. \ref{eq:cov} and a slight discrepancy between $\Var{M_1}$ and $\Var{M_2}$.  A full accounting of the effects of different types of Raman scattering events changes $\Condvar$ by less than $0.1$~dB---i.e. less than the least significant figure in our results---at the optimum photon numbers for conditional spin noise reduction and spin squeezing.

\subsection{Spin Noise in Rotated State}

In the inset to Fig.~1, we approximate the variance $\var{S_z}_\alpha$ of $S_z$ in a state that has been rotated by an angle $\alpha$ about $\aver{\vc{S}}$ between the squeezing measurement $M_1$ and readout measurement $M_2$ by
\begin{equation}
\var{S_z}_\alpha\approx\Var{M_1-M_2}\vert_\alpha-\Measvar,
\end{equation}
where $\Measvar = \Var{M_1-M_2}\vert_{\alpha=0}/2$.  This is a good approximation for all $\alpha$ in the limit where $\Measvar\ll\CSSvar$ and $\Delta S_\mathrm{max}\gg\Delta S_\mathrm{min}$.

\subsection{Contrast} \label{sec:contrast}
\begin{figure}
  \includegraphics[width=\columnwidth]{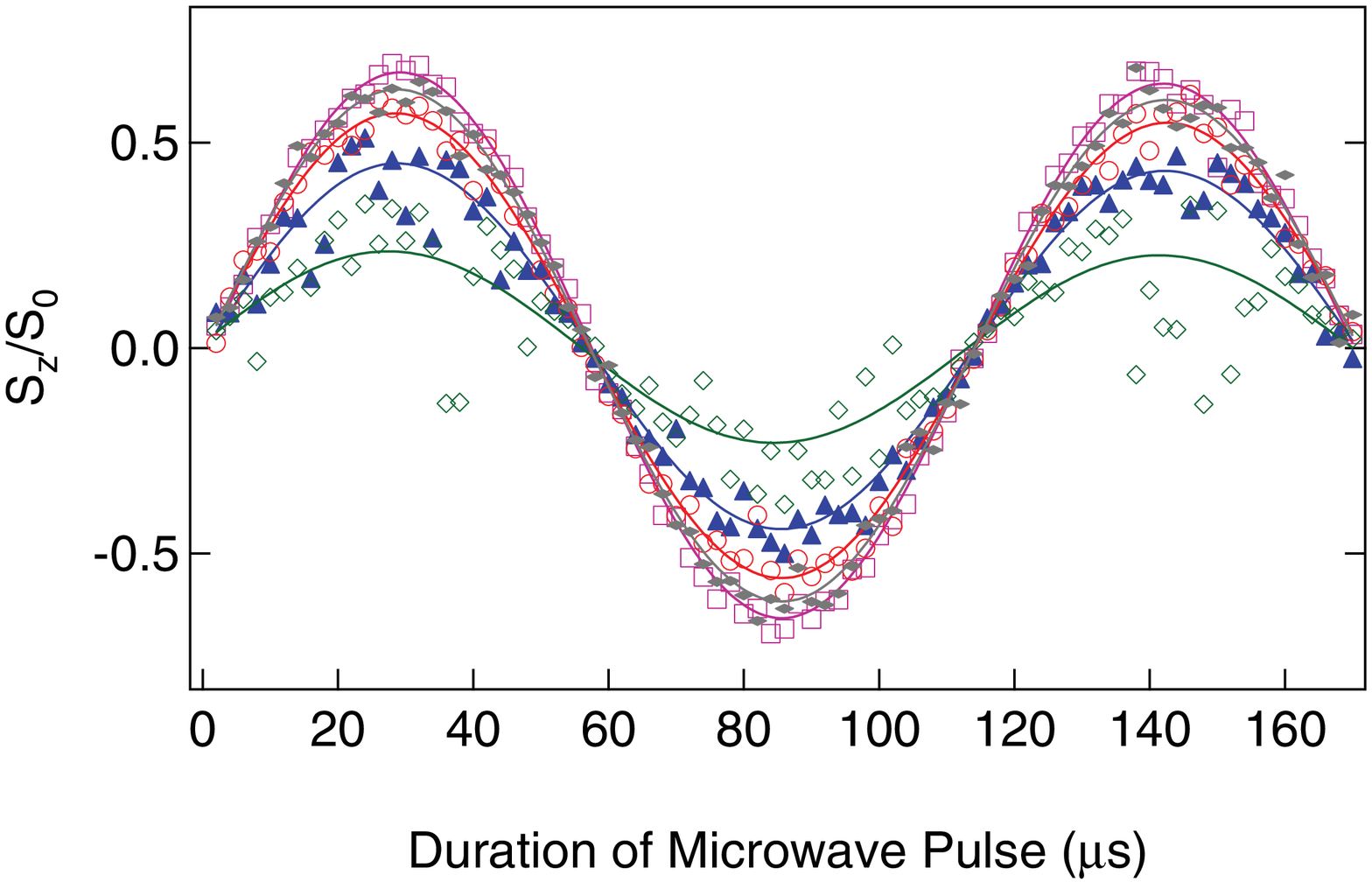}
  \caption{\label{fig:rabi} Measurement of clock contrast. The open squares correspond to the initial contrast with only lock light, the other curves to probe photon numbers between $p=10^5$ and $p=9\times 10^5$.}
\end{figure}

To verify and quantify spin squeezing (Fig.~3), we measure the contrast of a Rabi oscillation after the application of probe light (Fig.~\ref{fig:rabi}) using an atom number $N_0 = 4.0(1)\times 10^3$.  The Rabi oscillation is driven by a microwave pulse of variable duration between the squeezing and readout measurements, during which time both the probe light and the resonator locking light are off.

We observe a contrast loss that is linear in probe photon number $p$, as well as a process that imparts shot-to-shot phase fluctuations (via imbalances in the intracavity probe power between the two spin echo pulses) and yields a reduction in $\meanS$ that is quadratic in $p$.  We therefore fit to the data in Fig.~3 the expression $\mathcal{C} = \mathcal{C}_0 \exp(-\alpha p-\beta p^2/2)$, obtaining $\alpha = 7(1)\times 10^{-7}$, $\beta = 9(4)\times10^{-13}$, and $\mathcal{C}_\mathrm{0}=0.69(1)$.

As discussed in Sec. \ref{sec:metrologyfactor}, we define $\mathcal{C}_\mathrm{in}$ to be the contrast in the ensemble without squeezing ($p$=0) as observed in an ideal readout which flips no spins.  By comparing the two probe pulses constituting our real readout measurement, which uses a total of $5\times10^5$ probe photons, we determine that the reduction in observed contrast due to photon scattering or imperfect microwave rotations during the readout is at most 4(2)\%.  Correcting for this effect, we obtain $\mathcal{C}_\mathrm{in}=0.71(2)$.

The contrast measurement is performed at lower atom number than the noise measurements in Fig.~3 for two reasons.  First, at large atom number, atom projection noise augments the imbalances in intracavity probe power between the spin echo pulses and thus augments the resulting phase fluctuations; this effect can in principle be compensated using the result of the squeezing measurement, but we have not yet done so.  Second, at lower atom number, the entire Rabi oscillation curve is in the linear regime of the Lorentzian resonator transmission profile.  We verify that the contrast is independent of atom number by also measuring at $N_0=3.5(3)\times 10^4$ the portions of each Rabi oscillation curve that lie in the linear regime of the Lorentzian.  A fit to these portions of the curve taken with $p=0$ (i.e., with lock light only) yields $\mathcal{C}_\mathrm{in}=0.63(6)$.  While this measurement incurs greater uncertainty than that at small atom number, it confirms that the contrast is invariant across an order of magnitude in atom number.

\section{Fundamental Limits}

We outline here a derivation of the fundamental limit on squeezing in our system \cite{Madsen04,Hammerer04}.  A more complete treatment is given by Madsen and M\o{}lmer \cite{Madsen04}.

An ideal, Heisenberg-area-preserving measurement which adds no noise (i.e. a measurement where the product of squeezing and antisqueezing is unity) can reduce the variance of $N=2S_z$ by an amount inversely proportional to the broadening it imparts in phase.  For our dispersive optical measurement, this broadening comes from the photon shot noise uncertainty on the AC Stark shift due to the probing light.  For a given transmitted probe photon number $p$, the maximum photon-shot-noise-induced phase broadening is achieved by probing on cavity resonance and allows the normalized $S_z$ noise to be reduced to
\begin{equation}\label{eq:idealmeas}
\sigma^2\equiv\frac{\Condvar}{\CSSvar}=\frac{1}{1+N_0 p\phieff^2},
\end{equation}
where $\phieff = \phi_0\eta_\mathrm{eff}/\eta_0$ is the effective phase shift per transmitted photon (see Sec. \ref{sec:PhotonPhase}).  However, our real measurement adds noise because photon scattering can flip the atomic pseudo-spins.  Let $\Psc$ denote the probability for an atom in a superposition of states $\ket{1}$ and $\ket{2}$ to scatter a photon per probe photon transmitted through the resonator.  Then in the large-detuning limit $\delta\gg\Gamma$,
\begin{align}
\phieff &= 2\frac{\delta}{\Gamma}\Psc\nonumber\\
\Rightarrow N_0\phieff^2/\Psc &= N_0 \left(\frac{\delta}{\Gamma/2}\right)^2 \Psc = 2\Coop,
\end{align}
since $(2\delta/\Gamma)^2 \Psc = 2\eta_\mathrm{eff}$ is the single-atom resonant optical depth (see Sec. \ref{sec:coop}).  Thus, Eq. \ref{eq:idealmeas} can be expressed in differential form as $d\sigma^2/dp = -2\Coop\Psc(\sigma^2)^2$.  The only scattering events which contribute noise are Raman scattering events occurring with probability $\PRam$ per transmitted probe photon.  We assume, for simplicity, the worst-case scenario that each Raman-scattered photon transfers an atom from one clock state to the other.  Adding the noise contribution from these spin flips and neglecting absorption (which for our largest atom number $N_0=33000$ is $0.6\%$), one obtains \cite{Madsen04}
\begin{equation}
\frac{d\sigma^2}{dp} = -2\Coop\Psc(\sigma^2)^2 + 4\PRam.
\end{equation}
For $\Coop\gg 1$, the normalized spin noise $\sigma^2$ has a minimum of
\begin{equation}\label{eq:zetamin}
\sigma^2_\mathrm{min} = \sqrt{\frac{2}{\Coop}\frac{\PRam}{\Psc}}.
\end{equation}
In our system $\Psc=3.0\PRam$.  Thus, for our largest atom number $N_0=3.3\times 10^4$ and our probe polarization and detuning, where $N_0\eta_\mathrm{eff}=3100$, the minimum achievable normalized spin noise is $\sigma^2_\mathrm{min}=-18$~dB.

So far, we have neglected the fundamental contrast loss due to scattering.  Including this effect \cite{Madsen04}, one finds that $\sigma^2_\mathrm{min}$ is reached at $p\PRam=\frac{\sigma^2_\mathrm{min}}{8} \ln(\frac{8}{\sigma^2_\mathrm{min}})=0.012$ for our system.  The associated contrast reduction is only $1-\mathcal{C}\approx p(\frac{P_\mathrm{Ray,1}+P_\mathrm{Ray,2}}{2}-\sqrt{P_\mathrm{Ray,1}P_\mathrm{Ray,2}}+P_\mathrm{Ram})=0.012$, where $P_{\mathrm{Ray},F}$ is the Rayleigh scattering probability in state $\ket{F}$ \cite{Ozeri05}.  Thus, one also obtains a metrological squeezing parameter $\zeta_\mathrm{m,min}=\sigma^2_\mathrm{min}/\mathcal{C}^2=-18$~dB.  Reaching this fundamental limit would require detection of all the information leaving the resonator (e.g. by performing a phase measurement on resonance \cite{Teper08}) with perfect quantum efficiency and photon-shot-noise-limited sensitivity.

\end{bibunit}

\end{document}